# Status of the Chronopixel Project

C.Baltay[1], J.Brau[2], W.Emmet[1], D.Rabinovich[1], N.Sinev[2] and D.Strom[2] [*]

1 – Department of Physics, Yale University
New Haven, CT 06520-8120, USA

2 – Department of Physics, University of Oregon
Eugene, OR 97403-1274, USA

Other the past few years we have developed a monolithic CMOS pixel detector design for the ILC in collaboration with the SARNOFF Corporation. The unique feature of this design is the recorded time tag for each hit, allowing assignment of the hit to a particular bunch crossing (thus the name Chronopixel). The prototype design was completed in 2007. The first set of prototype devices was fabricated in 2008. We have developed a detailed testing plan and have designed the test electronics in collaboration with SLAC. Testing is expected to start early in 2009.

## 1  Introduction

The ILC environment offers a unique opportunity to achieve exceptional physics goals due to modest event rates, relative rates of background to signal and relatively low radiation levels. Precision measurement of the branching ratios for many of the Higgs decay modes is a primary goal. For this superb impact parameter resolution is needed. Pixilated vertex detectors are capable of exceptional performance. The CCD vertex detector of SLD [2] achieved 3.9 microns space point resolution. An advanced pixilated multi-layer vertex detector is planned for the linear collider.

The time structure of the ILC necessitates an extremely fast readout of the vertex detector. Monolithic CMOS pixel detectors allow extremely fast non-sequential readout of only those pixels that have hits in them. Recognizing its potential, we initiated an R&D effort to develop such a device [3]. Another important feature of our present conceptual design for these CMOS detectors is the possibility of putting a time stamp on each hit with sufficient precision to assign each hit to a particular bunch crossing. This significantly reduces the effective backgrounds because in the reconstruction of any particular event of interest we only need to consider those hits in the vertex detector that came from the same or adjacent bunch crossing.

## 2  Current status of the project

### 2.1  First prototype

Detailed design of the first prototype of the Monolithic CMOS detector, called Chronopixel, was completed by SARNOFF in January 2007. The main goal for the design and manufacture of these prototype devices was the proof of principle. To eventually achieve the ultimate design parameters of pixel size of about 10x10 $\mu m^2$ with 99% charged particle registration efficiency, advanced manufacturing technologies are required with feature size 45 nm, high resistivity (few K$\Omega$•cm), thick epitaxial layer (~15 $\mu$m) and deep p-well. This technology is still extremely expensive. We decided to use the standard TSMC process with 180 nm feature size, low resistivity (~10 Ω•cm), thin (7 μm) epitaxial layer, and only deep n-wells for the initial prototype. This choice leads to the minimum pixel size 50x50 μm$^2$ and very poor efficiency. But it permits a check on the main assumptions of noise level, power consumption and digital circuitry operations.

[*] This work is supported by DOE funding.

A batch of the first prototype devices (80 chips, each containing 80x80 array of 50x50 μm$^2$ pixels) was manufactured in 2008. 40 devices were packaged. As seen on figure 1, the package top cover is attached with soft glue, permitting its removal for testing with Fe55 source or IR laser.

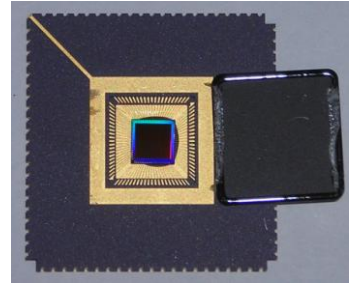

**Figure 1:** Photograph of the first prototype device

Two versions of the pixels were implemented in this prototype – pixel A and Pixel B. In pixel B all CMOS electronics is encapsulated in deep n-well. This insulates p-type doped regions from p-type epitaxial layer, thus allowing application of the negative bias to substrate, which increases depleted depth.

### 2.2 Design description

The architecture of each pixel of the first prototype is shown in Figure 2. After each bunch crossing, the signal in each pixel is compared to a preset, calibrated threshold level. If the signal is above threshold, the time of the bunch crossing is stored in the first slot of the 14 bit, 2 deep timing memory array. Time of an exceeding threshold hit from a subsequent bunch crossing is stored in the second memory slot. These time stamps are obtained by making snapshots of the 14 bit bus, controlled by an external clock. During the 200 ms gap between bunch trains all recorded timestamps from pixels with hits are read out, and the pixel memory is reset.

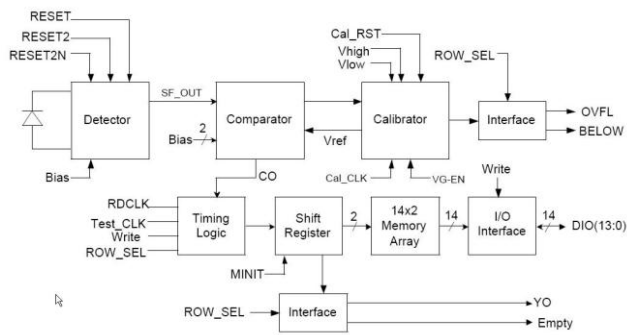

**Figure 2:** One pixel block-diagram

The critical element of this design is the calibrator. Calibration of the thresholds is performed during the 200 ms interval between bunches. To calibrate thresholds against individual comparator offsets, the reference voltage Vref is ramped in eight 0.125 mV steps. The step number at which the comparator fires is recorded in the calibration register, and the corresponding voltage offset is added to the comparator threshold for the given pixel.

The design also incorporates a special "soft reset" circuit, a proprietary technique of the SARNOFF Corporation, which is claimed to reduce σ of noise to 25 e .

### 2.3 Test plan

Special electronics for Chronopixel prototype testing is been developed by SLAC [4]. All necessary power supplies and bias voltages, control signals and monitors of currents, voltages and package temperature will be provided. We plan to do following tests:
  A. Calibration test to check that all pixels fall within 2 mV calibration window;
  B. Memory test of recording and read back from pixel memory;
  C. Fe55 source test to measure pixel sensitivity and noise;
  D. IR laser tests to check proper time stamping.

### 2.4 Expected performance

We simulated the expected signals from the prototype device. The electric field map is obtained from the TCAD device simulation, and the charge propagation inside the silicon sensors is simulated by software of N.Sinev, developed for ILC detector simulation. Figure 3 shows the expected distribution of signal amplitudes for Fe55 hits.

Most hits are generated outside the area of charge collection on the signal electrode. This is the reason for the large peak around zero. However, the small peak on the right corresponds to fully collected charge from the hit, and its position indicates pixel sensitivity. The noise level can be obtained from the width of the peak. As mentioned above, we expect the noise level at about 25 e σ.

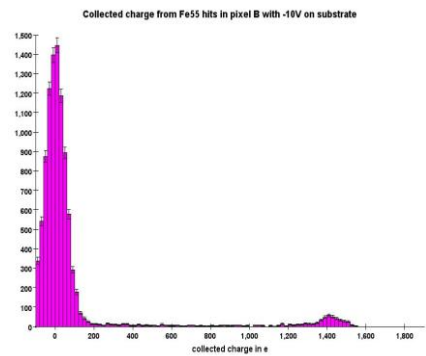

**Figure 3:** Amplitude distribution of Fe55 signals in pixel B

## 3 Future development

As soon as the design of the first prototype is validated, the second prototype design will start. All problems discovered in prototype one will be fixed, and the device will be designed to achieve highly efficient track detection. For this, deep p-well technology will be employed as well as a high resistivity thick epitaxial layer. The pixel size may still be about 50x50 μm. The scaling to smaller pixels will be the goal of the third prototype, which will be close to ultimate device. We expect the second prototype will be developed by the end of 2009 and the third by the end of 2010.